\documentclass[11pt,leqno]{article}

\usepackage{mathptmx}

\usepackage{amssymb,amsmath,amsthm}
\usepackage{comment}
\usepackage[latin1]{inputenc}    
\usepackage{pifont}    

\makeatletter%

\usepackage{nicefrac}
\usepackage{mathptmx}
\newcommand{\doublespacing}{\let\CS=
\@currsize\renewcommand{\baselinestretch}{1.75}\tiny\CS}
\newcommand{\extradoublespacing}{\let\CS=
\@currsize\renewcommand{\baselinestretch}{1.9}\tiny\CS}
\newcommand{\draftspacing}{\let\CS=
\@currsize\renewcommand{\baselinestretch}{2.0}\tiny\CS}
\newcommand{\hugedraftspacing}{\let\CS=
\@currsize\renewcommand{\baselinestretch}{2.4}\tiny\CS}
\makeatother%

\newcommand{\OMIT}[1]{} %
\newcommand{\gfootnote}[1]{} %

\newcommand\qedblob{\ding{113}}
\def\literalqed{{\ \nolinebreak\hfill\mbox{\qedblob\quad}}}

\newenvironment{proofs}{\noindent{\bf Proof.}\hspace*{1em}}{\literalqed\bigskip}

\newcommand{\seq}{\subseteq}

\newcommand{\scoresublevel}[3]{\mathit{score}_{#1}^{#2}(#3)}

\newcommand{\xthreec}{\mbox{\sc X3C}}
\newcommand{\restrictedhittingset}{\mbox{\sc Restricted Hitting Set}}

\newcommand{\p}{\mbox{\rm P}}

\newcommand{\np}{\mbox{\rm NP}}

\newcommand{\condition}{\,|\:}

\newenvironment{desctight}
  {\begin{list}{}{\setlength\labelwidth{0pt}%
        \setlength{\itemsep}{0.5pt}%
        \setlength{\parsep}{0pt}%
        \setlength\itemindent{-\leftmargin}%
        }}
    {\end{list}}

  \newtheorem{theorem}{Theorem}[section]
  \newtheorem{corollary}[theorem]{Corollary}

  \newtheorem{lemma}[theorem]{Lemma}

  \newtheorem{proposition}[theorem]{Proposition}

  \newtheorem{construction}[theorem]{Construction}
\usepackage{ulem}
\begin{document}

\title{Bucklin Voting is Broadly Resistant to Control\thanks{Supported
in part by the DFG under grants
RO~\mbox{1202/12-1} (within the European Science
Foundation's EUROCORES program LogICCC:
``Computational Foundations of Social Choice'') and RO~\mbox{1202/11-1}.
  Work done in part while the first and third
  author were visiting NICTA, Sydney, and the University of Newcastle.
  Some results for Bucklin voting in the present paper supersede the
  corresponding result for fallback voting shown 
  in~\cite{erd-pir-rot:t-With-Complete-CATS10-Ptr:fallback-voting}.}}

\author{G\'{a}bor Erd\'{e}lyi\thanks{URL: 
\mbox{\tt{}ccc.cs.uni-duesseldorf.de/\mbox{\tiny$\sim\,$}erdelyi.}
}, \ 
Lena Piras,
\ and \ 
J\"{o}rg Rothe\thanks{URL: 
\mbox{\tt{}ccc.cs.uni-duesseldorf.de/\mbox{\tiny$\sim\,$}rothe.}
} \\
Institut f\"{u}r Informatik \\
Heinrich-Heine-Universit\"{a}t D\"{u}sseldorf \\
40225 D\"{u}sseldorf \\
Germany
}

\date{May 22, 2010}

\maketitle

\begin{abstract}
  Electoral control models ways of changing the outcome of an election
  via such actions as adding/deleting/partitioning either candidates
  or voters.  These actions modify an election's participation
  structure and aim at either making a favorite candidate win
  (``constructive control'') or prevent a despised candidate from
  winning (``destructive control''), which yields a total of 22
  standard control scenarios.  To protect elections from such control
  attempts, computational complexity has been used to show that
  electoral control, though not impossible, is computationally
  prohibitive.
  Among natural voting systems with a polynomial-time winner problem,
  the two systems with the highest number of proven resistances to
  control types (namely 19 out of
  22~\cite{erd-now-rot:j:sp-av,erd-rot:c:fallback-voting,erd-pir-rot:t-With-Complete-CATS10-Ptr:fallback-voting})
  are ``sincere-strategy preference-based approval voting'' (SP-AV, a
  modification~\cite{erd-now-rot:j:sp-av} of a system proposed by
  Brams and
  Sanver~\cite{bra-san:j:critical-strategies-under-approval}) and
  fallback voting~\cite{bra-san:j:preference-approval-voting}.  Both
  are hybrid systems; e.g., fallback voting combines approval with
  Bucklin voting.
  In this paper, we study the control complexity of Bucklin voting
  itself and show that it behaves equally well in terms of control
  resistance for the 20 cases investigated so far.  As Bucklin voting
  is a special case of fallback voting, all resistances shown for
  Bucklin voting in this paper strengthen the corresponding resistance
  for fallback voting shown
  in~\cite{erd-pir-rot:t-With-Complete-CATS10-Ptr:fallback-voting}.
\end{abstract}
\maketitle                   %

\section{Introduction}
 \label{sec:introduction}

Since the seminal paper of Bartholdi et
al.~\cite{bar-tov-tri:j:control}, the complexity of \emph{electoral
  con\-trol}---changing the outcome of an election via such actions as
adding/de\-le\-ting/partitioning either candidates or voters---has
been studied for a variety of voting systems.  Unlike
\emph{manipulation}~\cite{bar-tov-tri:j:manipulating,bar-orl:j:polsci:strategic-voting,con-san-lan:j:when-hard-to-manipulate,fal-hem-hem-rot:j:llull-copeland-full-techreport},
which models attempts of strategic voters to influence the outcome of
an election via casting insincere votes, control models ways of an
external actor, the ``chair,'' to tamper with an election's
participation structure so as to alter its outcome.  Another way of
tampering with the outcome of elections is
\emph{bribery}~\cite{fal-hem-hem:j:bribery,fal-hem-hem-rot:j:llull-copeland-full-techreport},
which shares with manipulation the feature that votes are being
changed, and with control the aspect that an external actor tries to
change the outcome of the election.  Faliszewski et
al.~\cite{fal-hem-hem-rot:b:richer} survey known complexity results
for control, manipulation, and bribery.

Elections have been used for preference aggregation not only in the
context of politics and human societies, but also in artificial
intelligence, especially in multiagent systems, and other topics in
computer science (see, e.g.,
\cite{eph-ros:j:multiagent-planning,gho-mun-her-sen:c:voting-for-movies,dwo-kum-nao-siv:c:rank-aggregation}).
That is why it is important to study the computational properties of
voting systems.  In particular, complexity can be used to protect
elections against tampering attempts in control, manipulation, and
bribery attacks by showing that such attacks, though not impossible in
principle, are computationally prohibitive.

Regarding control, a central question is to find voting systems that
are computationally resistant to as many of the common $22$ control
types as possible, where resistance means the corresponding control
problem is $\np$-hard.  Each control type is either constructive (the
chair seeking to make some candidate win) or destructive (the chair
seeking to make some candidate not end up winning).
Erd\'{e}lyi and Rothe~\cite{erd-rot:c:fallback-voting} proved
that fallback voting~\cite{bra-san:j:preference-approval-voting}, a
hybrid voting system combining Bucklin with approval voting, is
resistant to each of these $22$ standard control types except five
types of voter control.  They proved that fallback voting is
vulnerable to two of those control types (i.e., these control problems
are polynomial-time solvable), leaving the other three cases open.
Erd\'{e}lyi, Piras and Rothe~\cite{erd-pir-rot:t-With-Complete-CATS10-Ptr:fallback-voting}
recently proved that fallback voting is resistant to constructive and
destructive control by partition of voters in the tie-handling model
``ties promote.''

Thus fallback voting is not only fully resistant to candidate
control~\cite{erd-rot:c:fallback-voting} but also fully resistant to
constructive control.  In terms of the total number of proven
resistances it draws level with ``sincere-strategy preference-based
approval voting'' (SP-AV,
another hybrid system proposed by Brams and
Sanver~\cite{bra-san:j:critical-strategies-under-approval}): Both
have the most (19 out of 22) proven resistances to control among
natural voting systems with a polynomial-time winner problem.  Among
such systems, only plurality and SP-AV were previously known to be
fully resistant to candidate control
\cite{bar-tov-tri:j:control,hem-hem-rot:j:destructive-control,erd-now-rot:j:sp-av},
and only Copeland voting and SP-AV were previously known to be fully
resistant to constructive control
\cite{fal-hem-hem-rot:j:llull-copeland-full-techreport,erd-now-rot:j:sp-av}.
However, plurality has fewer resistances to voter control,
Copeland voting has fewer resistances to destructive control,
and SP-AV is arguably less natural a system than fallback
voting~\cite{erd-now-rot:j:sp-av,bau-erd-hem-hem-rot:t:computational-apects-of-approval-voting}.

Even more natural than fallback voting, however, is Bucklin voting
itself, one of its two constituent systems.  After all, fallback
voting is a hybrid system in which voters are required to provide two
types of preference (a ranking of candidates and an approval vector),
whereas in Bucklin voting it is enough for the voters to rank the
candidates.  Moreover, Bucklin voting has the important property that
it is \emph{majority-consistent}, which means that whenever a majority
candidate exists,\footnote{A \emph{majority candidate} is a candidate
  that is ranked at top position in more than half of the votes.}  he
or she is the (unique) Bucklin winner.  In contrast, fallback voting
is not majority-consistent.

In this paper, we study the control complexity of Bucklin voting and
show that it has as many control resistances as fallback voting in the
20 cases investigated so far.  In particular, Bucklin voting is also
fully resistant to both candidate control and constructive control.

This paper is organized as follows.  In
Section~\ref{sec:preliminaries}, some basic notions from social choice
theory, and in particular Bucklin voting, as well as the 22 standard 
types of control are defined.  Our results on the control
complexity of Bucklin voting are presented in
Section~\ref{sec:results}.  Finally, Section~\ref{sec:conclusions}
provides some conclusions and open questions.

\section{Preliminaries}
\label{sec:preliminaries}

\subsection{Elections and Voting Systems}

An \emph{election} $(C,V)$ is given by a finite set $C$ of candidates
and a finite list $V$ of votes over~$C$.  A \emph{voting system} is
a rule that specifies how to determine the winner(s) of any given
election.  The two voting systems considered in this paper are Bucklin
voting and fallback voting.

In \emph{Bucklin voting}, votes are represented as linear orders
over~$C$, i.e., each voter ranks all candidates according to his or
her preferences.  For example, if $C = \{a, b, c, d\}$ then a vote
might look like $\begin{array}{@{\, }c@{\ \ }c@{\ \ }c@{\ \ }c} c & d
  & a & b,
\end{array}$
i.e., this voter (strictly) prefers $c$ to~$d$, $d$ to~$a$,
and $a$ to~$b$.  Given an election $(C,V)$ and a candidate $c \in C$,
define the \emph{level~$i$ score of $c$ in $(C,V)$} (denoted by
$\scoresublevel{(C,V)}{i}{c}$) as the number of votes in $V$ that rank
$c$ among their top $i$ positions.  Denoting the \emph{strict majority
  threshold for a list $V$ of voters} by $\mbox{maj}(V) = \lfloor
\nicefrac{\|V\|}{2}\rfloor +1$, the \emph{Bucklin score of $c$ in
  $(C,V)$} is the smallest $i$ such that $\scoresublevel{(C,V)}{i}{c}
\geq \mbox{maj}(V)$.  All candidates with a smallest Bucklin score,
say~$k$, and a largest level~$k$ score are the \emph{Bucklin winners 
(BV winners, for short) in $(C,V)$}.
If some candidate becomes a Bucklin winner on level~$k$,
we call him or her a \emph{level~$k$ BV winner in $(C,V)$}.  Note that
a level~$1$ BV winner must be unique, but there may be more level~$k$
BV winners than one for $k>1$, i.e., an election may have more than
one Bucklin winner in general.

As a notation, when a vote contains a subset of the candidate set,
such as 
$\begin{array}{@{}c@{\ \ }c@{\ \ }c@{}}
 c & D & a
\end{array}$ 
for a subset $D \subseteq C$, this is a shorthand for 
$\begin{array}{@{}c@{\ \ }c@{\ \ }c@{\ \ }c@{\ \ }c@{}}
 c & d_1 & \cdots & d_{\ell} & a
\end{array}$,
where the elements of $D = \{d_1, \ldots , d_{\ell}\}$ are ranked with
respect to some (tacitly assumed) fixed ordering of all candidates in~$C$.
For example, if $C = \{a, b, c, d\}$ and $D = \{b, d\}$ then
``$\begin{array}{@{}c@{\ \ }c@{\ \ }c@{}}
 c & D & a
\end{array}$''
is a shorthand for the vote
$\begin{array}{@{}c@{\ \ }c@{\ \ }c@{\ \ }c@{}}
 c & b & d & a
\end{array}$.

\subsection{Types of Electoral Control}

There are $11$ types of electoral control, each coming in two
variants.  In \emph{constructive
  control}~\cite{bar-tov-tri:j:control}, the chair tries to make his
or her favorite candidate win; in \emph{destructive
  control}~\cite{hem-hem-rot:j:destructive-control}, the chair tries
to prevent a despised candidate's victory.  We refrain from giving a
detailed discussion of real-life scenarios for each of these $22$
standard control types that motivate them; these can be found in,
e.g.,
\cite{bar-tov-tri:j:control,hem-hem-rot:j:destructive-control,fal-hem-hem-rot:j:llull-copeland-full-techreport,hem-hem-rot:j:hybrid,erd-now-rot:j:sp-av}.
However, we stress that every control type is motivated by an
appropriate real-life scenario.

We start with
partition of voters with the tie-handling rule ``ties promote'' (TP),
see Hemaspaandra et al.~\cite{hem-hem-rot:j:destructive-control}.
This control type produces a two-stage election with two first-stage
and one final-stage subelections.  The constructive variant of this
problem is:
\begin{desctight}

\item[Name:] {\sc Constructive Control by Partition of Voters in TP}.
\item[Instance:] A set $C$ of candidates, a list $V$ of votes
  over~$C$, and a designated candidate $c \in C$.
\item[Question:] Can $V$ be partitioned into $V_1$ and $V_2$ such that
  $c$ is the unique winner of the two-stage election in which the
  winners of the two first-stage subelections, $(C,V_1)$ and
  $(C,V_2)$, run against each other in the final stage?

\end{desctight}

The destructive variant of this problem is defined analogously, except
it asks whether $c$ is \emph{not} a unique winner of this two-stage
election.  In both variants, if one uses the tie-handling model TE
(``ties eliminate,'' see~\cite{hem-hem-rot:j:destructive-control})
instead of TP in the two first-stage subelections, a winner $w$ of
$(C,V_1)$ or $(C,V_2)$ proceeds to the final stage if and only if $w$
is the only winner of his or her subelection.  Each of the four
problems just defined models ``two-district gerrymandering.''

There are many ways of introducing new voters into an
election---think, e.g., of ``get-out-the-vote'' drives, or of lowering
the age-limit for the right to vote, or of attracting new voters with
certain promises or even small gifts), and such scenarios are modeled
as {\sc Constructive/Destructive Control by Adding Voters}: Given a
set $C$ of candidates, two disjoint lists of votes over $C$ (one list,
$V$, corresponding to the already registered voters and the other
list, $W$, corresponding to the as yet unregistered voters whose votes
may be added), a designated candidate $c \in C$, and a nonnegative
integer~$k$, is there a subset $W' \subseteq W$ such that $\|W'\| \leq
k$ and $c$ is (is not) the unique winner in $(C, V \cup W')$?

Disenfranchisement and other means of voter suppression is modeled as
{\sc Constructive/Destructive Control by Deleting Voters}: Given a set
$C$ of candidates, a list $V$ of votes over~$C$, a designated
candidate $c \in C$, and a nonnegative integer~$k$, can one make $c$
the unique winner (not a unique winner) of the election resulting from
deleting at most $k$ votes from~$V$?

Having defined these eight standard types of voter control, we now
turn to the $14$ types of candidate control.  Now, the control action
seeks to influence the outcome of an election by either adding,
deleting, or partitioning the candidates, again for both the
constructive and the destructive variant.

In the adding candidates cases, we distinguish between adding, from a
given pool of spoiler candidates, an \emph{unlimited} number of such
candidates (as
originally defined by Bartholdi et al.~\cite{bar-tov-tri:j:control})
and adding a \emph{limited} number of spoiler candidates (as defined
by Faliszewski et
al.~\cite{fal-hem-hem-rot:j:llull-copeland-full-techreport}, to stay
in sync with the problem format of control by deleting candidates and
by adding/deleting voters).  {\sc Constructive/Destructive Control by
  Adding (a Limited Number of) Candidates}, is defined as follows:
Given two disjoint candidate sets, $C$ and~$D$, a list $V$ of votes
over $C \cup D$, a designated candidate $c \in C$, and a nonnegative
integer~$k$, can one find a subset $D' \subseteq D$ such that $\|D'\|
\leq k$ and $c$ is (is not) the unique winner in $(C \cup D', V)$?
The ``unlimited'' version of the problem is the same, except that the
addition limit $k$ and the requirement ``$\|D'\| \leq k$'' are being
dropped, so \emph{any} subset of the spoiler candidates may be added.

{\sc Constructive/Destructive Control by Deleting Candidates} is
defined by: Given a set $C$ of candidates, a list $V$ of votes over
$C$, a designated candidate $c \in C$, and a nonnegative integer~$k$,
can one make $c$ the unique winner (not a unique winner) of the
election resulting from deleting at most $k$ candidates
(other than $c$ in the destructive case) from~$C$?

Finally, we define the partition-of-candidate cases, again using
either of the two tie-handling models, TP and TE, but now we define
these scenarios with and without a run-off.  The variant with run-off,
{\sc Constructive/Destructive Control by Run-Off Partition of
  Candidates}, is analogous to the par\-ti\-tion-of-voters control type:
Given a set $C$ of candidates, a list $V$ of votes over~$C$, and a
designated candidate $c \in C$, can $C$ be partitioned into $C_1$ and
$C_2$ such that $c$ is (is not) the unique winner of the two-stage
election in which the winners of the two first-stage subelections,
$(C_1,V)$ and $(C_2,V)$, who survive the tie-handling rule run against
each other in the final stage?  The variant without run-off is the
same, except that the winners of first-stage subelection $(C_1,V)$ who
survive the tie-handling rule run against $(C_2,V)$ in the final round
(and not against the winners of $(C_2,V)$ surviving the tie-handling
rule).\footnote{For example, think of a sports tournament in which
  certain teams (such as last year's champion and the team hosting
  this year's championship) are given an exemption from
  qualification.}

\subsection{Immunity, Susceptibility, Resistance, and Vulnerability}

Let $\mathfrak{CT}$ be a control type. We say a voting system is
\emph{immune to~$\mathfrak{CT}$} if it is impossible for the chair to
make the given candidate the unique winner in the constructive case
(not a unique winner in the destructive case) via exerting control of
type~$\mathfrak{CT}$.  We say a voting system is \emph{susceptible
  to~$\mathfrak{CT}$} if it is not immune to~$\mathfrak{CT}$.  A
voting system that is susceptible to~$\mathfrak{CT}$ is said to be
\emph{vulnerable to~$\mathfrak{CT}$} if the control problem
corresponding to $\mathfrak{CT}$ can be solved in polynomial time, and
is said to be \emph{resistant to~$\mathfrak{CT}$} if the control
problem corresponding to $\mathfrak{CT}$ is $\np$-hard.  These notions
are due to Bartholdi et al.~\cite{bar-tov-tri:j:control} (except that
we follow the now more common approach of Hemaspaandra et
al.~\cite{hem-hem-rot:j:hybrid} who define
\emph{resistant} to mean ``susceptible and $\np$-hard'' rather than
``susceptible and $\np$-complete'').

\section{Results}
\label{sec:results}

\subsection{Overview}
\label{sec:results:overview}

Table~\ref{tab:summary-of-results} shows in boldface our results on
the control complexity of Bucklin voting.  For comparison, this table
also shows the results for fallback voting that are due to Erd\'{e}lyi
et
al.~\cite{erd-rot:c:fallback-voting,erd-pir-rot:t-With-Complete-CATS10-Ptr:fallback-voting},
for approval voting that are due to Hemaspaandra et
al.~\cite{hem-hem-rot:j:destructive-control}, and for SP-AV that are
due to Erd\'{e}lyi et al.~\cite{erd-now-rot:j:sp-av}.

\begin{theorem}
\label{thm:Bucklin-summary-of-results}
Bucklin voting is resistant, vulnerable, and susceptible to the $22$
types of control defined in Section~\ref{sec:preliminaries} as shown
in Table~\ref{tab:summary-of-results}.
\end{theorem}

Since Bucklin voting is the special case of fallback voting where each
voter approves of every candidate, we have the following corollary.
Note that, by the first item of
Corollary~\ref{cor:Bucklin-summary-of-results}, the resistances for
Bucklin voting shown in the present paper imply all resistances for
fallback voting shown in
\cite{erd-rot:c:fallback-voting,erd-pir-rot:t-With-Complete-CATS10-Ptr:fallback-voting}
except one: the destructive case of partition of voters in the tie-handling model~{TP}.

\begin{corollary}
\label{cor:Bucklin-summary-of-results}
\begin{enumerate}
\item Fallback voting inherits all the $\np$-hardness lower bounds
  from Bucklin voting (i.e., if Bucklin voting is resistant to a
  control type $\mathfrak{CT}$ then fallback voting is also resistant
  to $\mathfrak{CT}$).
\item Bucklin voting inherits all the $\p$ membership upper bounds
  from fallback voting (i.e., if fallback voting is vulnerable to a
  control type $\mathfrak{CT}$ then Bucklin voting is also vulnerable
  to $\mathfrak{CT}$).
\end{enumerate}
\end{corollary}

\begin{table*}[t!]
\centering
{\footnotesize
\begin{tabular}{|l||l|l||l|l||l|l||l|l|}
\hline
                    & \multicolumn{2}{c||}{Bucklin Voting}
                    & \multicolumn{2}{c||}{Fallback Voting}
                    & \multicolumn{2}{c||}{SP-AV}
                    & \multicolumn{2}{c|}{Approval}
\\ \cline{2-9}
Control by          & Const. & Dest.
                    & Const. & Dest.
                    & Const. & Dest.
                    & Const. & Dest.
\\ \hline\hline
Adding Candidates (unlimited)
                    & {\bf R}               & {\bf R} 
                    & 	   R                &      R 
                    & 	   R                &      R 
                    &      I                &      V 
\\ \hline
Adding Candidates (limited)
                    & {\bf R}               & {\bf R}
                    & 	   R                &      R  
                    &      R                &      R 
                    &      I                &      V 
\\ \hline
Deleting Candidates & {\bf R}               & {\bf R}
                    & 	   R                &      R  
                    &      R                &      R 
                    &      V                &      I 
\\ \hline
Partition of Candidates
                    & {\bf TE: R}           & {\bf TE: R}
                    &      TE: R            &      TE: R  
                    &      TE: R            &      TE: R 
                    &      TE: V            &      TE: I 
\\
                    & {\bf TP: R}           & {\bf TP: R} 
                    &      TP: R            &      TP: R  
                    &      TP: R            &      TP: R      
                    &      TP: I            &      TP: I 
\\ \hline
Run-off Partition of Candidates
                    & {\bf TE: R}           & {\bf TE: R}  
                    &      TE: R            &      TE: R  
                    &      TE: R            &      TE: R 
                    &      TE: V            &      TE: I 
\\ 
                    & {\bf TP: R}           & {\bf TP: R}
                    &      TP: R            &      TP: R   
                    &      TP: R            &      TP: R      
                    &      TP: I            &      TP: I 
\\ \hline
Adding Voters       & {\bf R}               & {\bf V}        
                    &      R                &      V          
                    &      R                &      V 
                    &      R                &      V 
\\ \hline
Deleting Voters     & {\bf R}               & {\bf V}      
                    &      R                &      V         
                    &      R                &      V 
                    &      R                &      V 
\\ \hline
Partition of Voters
                    &      {\bf TE: R}      &      {\bf TE: S}
                    &      TE: R            &      TE: S
                    &      TE: R            &      TE: V 
                    &      TE: R            &      TE: V 
\\
                    &      {\bf TP: R}      &      {\bf TP: S}
                    &      TP: R            &      TP: R  
                    &      TP: R            &      TP: R      
                    &      TP: R            &      TP: V 
\\ 
\hline
\end{tabular}
}
\caption{\label{tab:summary-of-results}
Overview of results and comparison with known results.
Key: 
I $=$ immune,
S $=$ susceptible,
R $=$ resistant,
V $=$ vulnerable,
TE $=$ ties eliminate, and TP $=$ ties promote.
Results new to this paper are in boldface.  
} 
\end{table*}

\subsection{Susceptibility}
\label{sec:results:susceptibility}

If an election system $\mathcal{E}$ satisfies the ``unique'' variant
of the Weak Axiom of Revealed Preference\footnote{This axiom says that
  the unique winner $w$ of any election is also the unique winner of every
  subelection including~$w$.} (Unique-WARP, for short), then
$\mathcal{E}$ is immune to constructive control by adding candidates
(by adding a limited number of candidates), and this observation has
been applied to approval
voting~\cite{bar-tov-tri:j:control,hem-hem-rot:j:destructive-control}.
Unlike approval voting but just as fallback voting, Bucklin voting does not
satisfy Unique-WARP.

\begin{proposition}
\label{prop:unique-warp}
Bucklin voting does not satisfy Unique-WARP.
\end{proposition}

\begin{proofs}
Consider the election $(C,V)$ with candidate set $C = \{a, b, c,
d\}$ and voter collection $V = \{v_1, v_2, \ldots , v_6\}$:
\[
\begin{array}{rc@{\ \ }c@{\ \ }c@{\ \ }c}
 & \multicolumn{4}{c}{(C,V)} \\ 
\cline{2-5}
v_1=v_2=v_3: & a & c & b & d \\
v_4=v_5:     & b & d & c & a \\
v_6:         & d & a & c & b \\
\end{array}
\]

Candidate $a$ is the unique Bucklin winner of the election $(C,V)$, 
reaching the strict majority threshold on level~2 
with $\scoresublevel{(C,V)}{2}{a}=4$.  By removing candidate $b$ from the
election, we get the subelection $(C',V)$ with $C'=\{a,c,d\}$.
There is no candidate on level~1 who passes the strict majority threshold.
However, there are two candidates on the second level with a strict
majority, namely candidates $a$ and~$c$.  Since
$\scoresublevel{(C',V)}{2}{c} = 5 > 4 = \scoresublevel{(C',V)}{2}{a}$,
the unique Bucklin winner of the subelection $(C',V)$ is
candidate~$c$. Thus, Bucklin voting does not satisfy
Unique-WARP.~\end{proofs}

Indeed, as we will now show, Bucklin voting is susceptible to each of
our $22$ control types.  Our proofs make use of the results of
\cite{hem-hem-rot:j:destructive-control} that provide general proofs
of and links between certain susceptibility cases.  For the sake of
self-containment, we state their results, as
Theorems~\ref{thm:voiced-control},
\ref{thm:duality-constructive-destructive-control},
and~\ref{thm:susceptibility-implications}, in the appendix.

We start with susceptibility to candidate control.

\begin{lemma}
 \label{lem:susceptible-candidate-control}
Bucklin voting is susceptible to constructive and destructive control
by adding candidates (in both the ``limited'' and the ``unlimited''
case), by deleting candidates, and by partition of candidates (with or
without run-off and for each in both model TE and model TP).
\end{lemma}

\begin{proofs}
From Theorem~\ref{thm:voiced-control} and the fact that Bucklin voting is a voiced
voting system,\footnote{An election system is said to be \emph{voiced}
  if the single candidate in any one-candidate election always wins.}
it follows that Bucklin voting is susceptible to constructive control by deleting
candidates, and to destructive control by adding candidates (in both
the ``limited'' and the ``unlimited'' case).

Now, consider the election $(C,V)$ given in the proof of
Proposition~\ref{prop:unique-warp}.  The unique Bucklin winner of the
election is candidate $a$.  Partition $C$ into $C_1=\{a,c,d\}$ and
$C_2=\{b\}$. The unique Bucklin winner of subelection $(C_1,V)$ is
candidate~$c$, as shown in the proof of
Proposition~\ref{prop:unique-warp}.  In both partition and run-off
partition of candidates and for each in both tie-handling models, TE
and TP, candidate $b$ runs against candidate $c$ in the final stage of
the election. The unique Bucklin winner is in each case candidate
$c$. Thus, Bucklin voting is susceptible to destructive control by partition of
candidates (with or without run-off and for each in both model TE and
model~TP).

By Theorem~\ref{thm:susceptibility-implications}, Bucklin voting is also
susceptible to destructive control by deleting candidates. By
Theorem~\ref{thm:duality-constructive-destructive-control}, Bucklin voting is also
susceptible to constructive control by adding candidates (in both the
``limited'' and the ``unlimited'' case).

Now, changing the roles of $a$ and $c$ makes $c$ our distinguished
candidate. In election $(C,V)$, $c$ loses against candidate $a$. By
partitioning the candidates as described above, $c$ becomes the unique
Bucklin winner of the election. Thus, Bucklin voting is susceptible to constructive
control by partition of candidates (with or without run-off and for
each in both tie-handling models, TE and TP).
\end{proofs}

We now turn to susceptibility to voter control.

\begin{lemma}
 \label{lem:susceptible-voter-control}
Bucklin voting is susceptible to constructive and destructive 
control by adding voters, by deleting voters, and by partition 
of voters (in both model TE and model TP).
\end{lemma}

\begin{proofs}
Consider the election $(C,V)$, where $C=\{a,b,c,d\}$ is the 
set of candidates and $V=\{v_1,v_2,v_3,v_4\}$ is the 
collection of voters with the following preferences:
\[
\begin{array}{lc@{\ \ }c@{\ \ }c@{\ \ }c}
 & \multicolumn{4}{c}{(C,V)} \\ \cline{2-5}
v_1: & a & c & b & d \\
v_2: & d & c & a & b \\
v_3: & b & a & c & d \\
v_4: & b & a & c & d \\
\end{array}
\]
We partition $V$ into $V_1=\{v_1,v_2\}$ and $V_2=\{v_3,v_4\}$. Thus we
split $(C,V)$ into two subelections:
\[
\begin{array}{lc@{\ \ }c@{\ \ }c@{\ \ }c@{\ \ }c@{\ \ }c@{\ \ }c@{\ \ }c@{\ \ }c}
 & \multicolumn{4}{c}{(C,V_1)} & \mbox{\quad and\quad }
 & \multicolumn{4}{c}{(C,V_2)}
\\ \cline{2-5}\cline{7-10}
v_1: & a & c & b & d &
     &   &   &   &   \\
v_2: & d & c & a & b & 
     &   &   &   &   \\
v_3: &   &   &   &   & 
     & b & a & c & d \\
v_4: &   &   &   &   & 
     & b & a & c & d \\
\end{array}
\]

Clearly, candidate $a$ is the unique Bucklin winner of $(C,V)$.
However, $c$ is the unique Bucklin winner of $(C,V_1)$ and $b$
is the unique Bucklin winner of $(C,V_2)$, and so $a$ is not 
promoted to the final stage. Thus, Bucklin voting is susceptible
to destructive control by partition of voters in
both tie-handling models, TE and TP.

By Theorem~\ref{thm:voiced-control} and the fact that Bucklin voting is a voiced
voting system, Bucklin voting is susceptible to destructive control by deleting
voters.  By
Theorem~\ref{thm:duality-constructive-destructive-control}, Bucklin voting is also
susceptible to constructive control by adding voters.

By changing the roles of $a$ and $c$ again, we can see that Bucklin voting is
susceptible to constructive control by partition of voters in both
model TE and model~{TP}. By
Theorem~\ref{thm:susceptibility-implications}, Bucklin voting is also susceptible
to constructive control by deleting voters.  Finally, again by
Theorem~\ref{thm:duality-constructive-destructive-control}, Bucklin voting is
susceptible to destructive control by adding voters.~\end{proofs}

\subsection{Candidate Control}
\label{sec:results:candidate-control}

Fallback voting is a hybrid system combining Bucklin voting with approval
voting.  While fallback and approval voting behave quite differently
with respect to immunity/vulnerability/resistance to control (contrast
the results of Hemaspaandra et al.\ on approval
voting~\cite{hem-hem-rot:j:destructive-control} with those of
Erd\'{e}lyi et al.~\cite{erd-rot:c:fallback-voting,erd-pir-rot:t-With-Complete-CATS10-Ptr:fallback-voting}
on fallback voting, see Table~\ref{tab:summary-of-results}),
Bucklin voting seems to behave equally well as
fallback voting in terms of control resistance.  In particular, like
fallback voting, Bucklin voting is also fully resistant to candidate control.

\begin{theorem}
 \label{thm:bv-candidate-control}
 Bucklin voting is resistant to each of the $14$ standard types of candidate
 control.
\end{theorem}

All reductions except one (namely that for constructive control by
deleting candidates, see
Lemma~\ref{lem:bv-deleting-candidates-constr}) apply
Construction~\ref{con:bv-resistance-general-candidate-control} below.  
This construction is based on that for fallback
voting~\cite{erd-rot:c:fallback-voting}; however, there are
significant differences.  In fallback voting, the disapproved
candidates need not be ranked and can
safely be ignored, since they cannot score points.  In Bucklin voting,
however, there are no disapproved candidates, so \emph{every}
candidate has to be placed at a suitable position in each vote to make
the reduction work.  Thus the reductions for Bucklin voting 
will be more specific
and the arguments more involved.  Also, since every candidate can
potentially score points in Bucklin voting, no matter what his or her
position in a vote is, we have to use a restricted version of
{\sc Hitting Set}
, which by Lemma~\ref{lem:rhs}
is also $\np$-complete:

\begin{desctight}

\item[Name:] $\restrictedhittingset$.

\item[Instance:] A set $B=\{b_1,b_2,\ldots , b_m\}$, a collection
  $\mathcal{S} =\{S_1,S_2,\ldots , S_n\}$ of nonempty subsets
  $S_i\subseteq B$ such that $n>m$, and a positive integer $k < m$.

\item[Question:] Does $\mathcal{S}$ have a hitting set of size at most
$k$, i.e., is there a set $B'\subseteq B$ with $\|B'\|\leq k$ such that
for each $i$, $S_i \cap  B'\neq \emptyset$?  
\end{desctight}

We first need to show that {\sc Restricted Hitting Set} is $\np$-complete
in order to apply 
Construction~\ref{con:bv-resistance-general-candidate-control} 
in the proof of Theorem~\ref{thm:bv-candidate-control}. 

\begin{lemma}
\label{lem:rhs}
{\sc Restricted Hitting Set} is $\np$-complete.
\end{lemma}

\begin{proofs}
It is immediate that
{\sc Restricted Hitting Set} is in $\np$.
To show $\np$-hardness, we reduce the (general) {\sc Hitting Set}
problem to {\sc Restricted Hitting Set}.
Given an instance $(\hat{B},\hat{\mathcal{S}},\hat{k})$ of
{\sc Hitting Set}
, where $\hat{B} = \{b_1, b_2,\ldots , b_{\hat{m}}\}$
is a set, $\hat{\mathcal{S}} = \{S_1, S_2,\ldots , S_{\hat{n}}\}$
is a collection of nonempty subsets of $\hat{B}$, and 
$\hat{k} \leq \hat{m}$ is a positive integer, define the following
instance $(B,\mathcal{S},k)$ of {\sc Restricted Hitting Set}:

\begin{eqnarray*}
B           & = & \left\{
\begin{array}{ll}
\hat{B} \cup \{a\} & \mbox{ if $\hat{n} \leq \hat{m}$} \\
\hat{B}            & \mbox{ if $\hat{n} > \hat{m}$}
\end{array}
\right.
\\
\mathcal{S} & = & \left\{
\begin{array}{ll}
\hat{\mathcal{S}} \cup \{S_{\hat{n}+1}, S_{\hat{n}+2}, \ldots,
  S_{\hat{m}+2}\} & \mbox{ if $\hat{n} \leq \hat{m}$} \\
\hat{\mathcal{S}} & \mbox{ if $\hat{n} > \hat{m}$}
\end{array}
\right.
\\
k & = & \left\{
\begin{array}{ll}
\hat{k} + 1 & \mbox{ if $\hat{n} \leq \hat{m}$} \\
\hat{k}     & \mbox{ if $\hat{n} > \hat{m}$,}
\end{array}
\right.
\end{eqnarray*}

where

\[
S_{\hat{n}+1} = S_{\hat{n}+2} = \cdots = S_{\hat{m}+2} = \{a\}.
\]

Let $n$ be the number of members of $\mathcal{S}$ and $m$ be the
number of elements of~$B$.  Note that if $\hat{n} > \hat{m}$ then
$(B,\mathcal{S},k) = (\hat{B},\hat{\mathcal{S}},\hat{k})$, so $n =
\hat{n} > \hat{m} = m$; and if $\hat{n} \leq \hat{m}$ then $n =
\hat{m}+2 > \hat{m}+1 = m$.  Thus, in both cases $(B,\mathcal{S},k)$
fulfills the restriction of {\sc Restricted Hitting Set}.

It is easy to see that $\hat{\mathcal{S}}$ has a hitting set of size
at most $\hat{k}$ if and only if $\mathcal{S}$ has a hitting set of
size at most $k$.  In particular, assuming $\hat{n} \leq \hat{m}$, if
$\hat{\mathcal{S}}$ has a hitting set $B'$ of size at most $\hat{k}$
then $B' \cup \{a\}$ is a hitting set of size at most $k = \hat{k} +
1$ for $\mathcal{S}$; and if $\hat{\mathcal{S}}$ has no hitting set of
size at most $\hat{k}$ then $\mathcal{S}$ can have no hitting set of
size at most $k = \hat{k} + 1$ (because $a \not\in \hat{B}$, so $\{a\}
\cap S_i = \emptyset$ for each~$i$, $1 \leq i \leq
\hat{n}$).\end{proofs}

\begin{construction}
\label{con:bv-resistance-general-candidate-control}
Let $(B,\mathcal{S},k)$ be a given instance of
{\sc Restricted Hitting Set}, where $B = \{b_1, b_2, \ldots , b_m\}$
is a set, $\mathcal{S} = \{S_1, S_2, \ldots , S_n\}$ is a collection
of nonempty subsets $S_i \seq B$ such that $n>m$, and $k<m$ is a
positive integer.  (Thus, $n > m > k \geq 1$.)

Define the election $(C,V)$, where $C = B \cup \{c,d,w\}$ is the
candidate set and where $V$ consists of the following $6n(k+1)+4m+11$
voters:
\begin{center}
\begin{tabular}{|c||l|c|l|}
 \hline
$\#$ & For each \ldots & number of voters & ranking of candidates
 \\ \hline\hline
 1 & & $2m+1$ &
$
\begin{array}{c@{\ \ }c@{\ \ }c@{\ \ }c}
 c & d & B & w 
\end{array}
$ \\ \hline
 2 & & $2n+2k(n-1)+3$ &
$
\begin{array}{c@{\ \ }c@{\ \ }c@{\ \ }c}
 c & w & d & B
\end{array}
$ \\ \hline
 3 & & $2n(k+1)+5$ &
$
\begin{array}{c@{\ \ }c@{\ \ }c@{\ \ }c}
 w & c & d & B
\end{array}
$ \\ \hline
 4 & $i \in \{1, \ldots , n\}$ & $2(k+1)$ &
$
\begin{array}{c@{\ \ }c@{\ \ }c@{\ \ }c@{\ \ }c}
d & S_i & c & w & (B-S_i)
\end{array}
$ \\ \hline
 5 & $j \in \{1, \ldots , m\}$ & $2$ &
$
\begin{array}{c@{\ \ }c@{\ \ }c@{\ \ }c@{\ \ }c}
d & b_j & w & c & (B-\{b_j\})
\end{array}
$ \\ \hline
 6 & & $2(k+1)$ & 
$
\begin{array}{c@{\ \ }c@{\ \ }c@{\ \ }c}
 d & w & c & B
\end{array}
$ \\ \hline
\end{tabular}
\end{center}
\end{construction}

We now give a detailed proof of Theorem~\ref{thm:bv-candidate-control}
(except for the case of constructive control by deleting candidates,
which will be handled separately in
Lemma~\ref{lem:bv-deleting-candidates-constr}) via
Construction~\ref{con:bv-resistance-general-candidate-control} in
Lemmas~\ref{lem:bv-adding-candidates},
\ref{lem:bv-deleting-candidates-destr},
and~\ref{lem:bv-partition-candidates}.  The proofs of these lemmas in
turn will make use of Lemma~\ref{lem:bv-resistance-candidate-control}
below.

\begin{lemma}
\label{lem:bv-resistance-candidate-control}
Consider the election $(C,V)$ constructed according to
Construction~\ref{con:bv-resistance-general-candidate-control} from a
{\sc Restricted Hitting Set} instance $(B,\mathcal{S},k)$.
\begin{enumerate}
\item $c$ is the unique level~2 BV winner of $(\{ c,d,w \} ,V)$.

\item If $\mathcal{S}$ has a hitting set $B'$ of size~$k$, then $w$ is
the unique BV winner of election $(B' \cup \{c,d,w\},V)$.

\item Let $D \subseteq B \cup \{d,w\}$.  If $c$ is not a unique BV
  winner of election $(D \cup \{c\},V)$, then there exists a set $B'
  \seq B$ such that
\begin{enumerate}
\item \label{lem:bv-resistance-candidate-control-part-2a}
$D = B' \cup \{d,w\}$, 

\item \label{lem:bv-resistance-candidate-control-part-2b}
$w$ is the unique level~2 BV winner of election $(B' \cup
  \{c,d,w\},V)$, and

\item \label{lem:bv-resistance-candidate-control-part-2c}
$B'$ is a hitting set for $\mathcal{S}$ of size at most~$k$.
\end{enumerate}
\end{enumerate}
\end{lemma}

\begin{proofs}
  For the first part, note that there is no level~1 BV winner in
  election $(\{ c,d,w \} ,V)$ and we have the following level~2 scores
  in this election:
\begin{eqnarray*}
\scoresublevel{(\{ c,d,w \} ,V)}{2}{c} & = & 6n(k+1)+2(m-k)+9,  \\
\scoresublevel{(\{ c,d,w \} ,V)}{2}{d} & = & 2n(k+1)+4m+2k+3,  \mbox{ and} \\
\scoresublevel{(\{ c,d,w \} ,V)}{2}{w} & = & 4n(k+1)+2m+10.
\end{eqnarray*}

Since $n>m$ (which implies $n>k$), we have:
\begin{eqnarray*}
\scoresublevel{(\{ c,d,w \} ,V)}{2}{c}-\scoresublevel{(\{ c,d,w \} ,V)}{2}{d}
 & = & 4n(k+1)-(2m+4k)+6 > 0,\, \mbox{and}\\
\scoresublevel{(\{ c,d,w \} ,V)}{2}{c}-\scoresublevel{(\{ c,d,w \} ,V)}{2}{w}
 & = & 2n(k+1)-(2k+1) > 0.
\end{eqnarray*}
Thus, $c$ is the unique level~2 BV winner of $(\{ c,d,w \} ,V)$.

For the second part, suppose that $B'$ is a hitting set for
$\mathcal{S}$ of size~$k$.  Then there is no level~1 BV winner in
election $(B' \cup \{c,d,w\},V)$, and we have the following level~2
scores:
\begin{eqnarray*}
\scoresublevel{(B' \cup \{c,d,w\},V)}{2}{c}   &  =   & 4n(k+1)+2(m-k)+9,\\
\scoresublevel{(B' \cup \{c,d,w\},V)}{2}{d}   &  =   & 2n(k+1)+4m+2k+3,\\
\scoresublevel{(B' \cup \{c,d,w\},V)}{2}{w}   &  =   & 4n(k+1)+2(m-k)+10,\\
\scoresublevel{(B' \cup \{c,d,w\},V)}{2}{b_j} & \leq & 2n(k+1)+2
\mbox{\hspace{1ex} for all $b_j\in B'$}.
\end{eqnarray*}
It follows that $w$ is the unique level~2 BV winner of election $(B'
\cup \{c,d,w\},V)$.

For the third part, let $D \subseteq B \cup \{d,w\}$. Suppose $c$ is
not a unique BV winner of election $(D \cup \{c\},V)$.
\begin{enumerate}
\item[(\ref{lem:bv-resistance-candidate-control-part-2a})] Other
  than~$c$, only $w$ has a strict majority of votes on the
  second level and only $w$ can tie or beat $c$ in $(D \cup \{c\},V)$.
  Thus, since $c$ is not a unique BV winner of election $(D \cup
  \{c\},V)$, $w$ is clearly in~$D$.  In $(D\cup \{c\},V)$, candidate
  $w$ has no level~1 strict majority, and candidate $c$ has already on
  level~2 a strict majority.  Thus, $w$ must tie or beat $c$ on
  level~2.  For a contradiction, suppose $d\notin D$.  Then
\[
\scoresublevel{(D \cup \{c\},V)}{2}{c} \geq 4n(k+1)+2m+11.
\]
The level~2 score of $w$ is
\[
\scoresublevel{(D\cup \{c\},V)}{2}{w}=4n(k+1)+2m+10,
\]
which contradicts our assumption, that $w$ ties or beats $c$ on
level~2. Thus, $D=B' \cup \{d,w\}$, where $B'\subseteq B$.
   
\item[(\ref{lem:bv-resistance-candidate-control-part-2b})] This part
  follows immediately from
  part~(\ref{lem:bv-resistance-candidate-control-part-2a}).
   
\item[(\ref{lem:bv-resistance-candidate-control-part-2c})] Let $\ell$ be
  the number of sets in $\mathcal{S}$ not hit by $B'$.  We have that
\begin{eqnarray*}
\scoresublevel{(B' \cup \{c,d,w\},V)}{2}{w}
 & = & 4n(k+1)+10+2(m-\| B'\| ), \\
\scoresublevel{(B' \cup \{c,d,w\},V)}{2}{c}
 & = & 2(m-k)+4n(k+1)+9+2(k+1)\ell .
\end{eqnarray*}
From part~(\ref{lem:bv-resistance-candidate-control-part-2a}) we know
that
\[
\scoresublevel{(B' \cup \{c,d,w\},V)}{2}{w} \geq
\scoresublevel{(B' \cup \{c,d,w\},V)}{2}{c},
\]
so
\begin{eqnarray*}
4n(k+1)+10+2(m-\| B'\| )
 & \geq & 2(m-k)+4n(k+1)+9+2(k+1)\ell .
\end{eqnarray*}
The above inequality implies
\[
1 > \frac{1}{2} \geq \| B' \| -k + (k+1)\ell \geq 0,
\]
so $\| B' \| -k + (k+1)\ell = 0$.  Thus $\ell = 0$, and it follows
that $B'$ is a hitting set for $\mathcal{S}$ of size~$k$.
\end{enumerate}
This completes the proof of
Lemma~\ref{lem:bv-resistance-candidate-control}.~\end{proofs}

\begin{lemma}
\label{lem:bv-adding-candidates}
Bucklin voting is resistant to constructive and destructive control
by adding candidates (both in the limited and the unlimited version of
the problem).
\end{lemma}

\begin{proofs}
  Susceptibility holds by
  Lemma~\ref{lem:susceptible-candidate-control}.  $\np$-hardness
  follows immediately from Lemmas~\ref{lem:rhs}
  and~\ref{lem:bv-resistance-candidate-control}, via mapping the
{\sc Restricted Hitting Set} instance $(B,\mathcal{S},k)$ to the instance
\begin{enumerate}
\item $((\{c,d,w\} \cup B, V), w, k)$ of Constructive Control by
  Adding a Limited Number of Candidates,
\item $((\{c,d,w\} \cup B, V), c, k)$ of Destructive Control by Adding
  a Limited Number of Candidates,
\item $((\{c,d,w\} \cup B, V), w)$ of Constructive Control by
  Adding an Unlimited Number of Candidates, and
\item $((\{c,d,w\} \cup B, V), c)$ of Destructive Control by Adding
  an Unlimited Number of Candidates.
\end{enumerate}
where in each case $c$, $d$, and $w$ are the qualified candidates and
$B$ is the set of spoiler candidates.~\end{proofs}

\begin{lemma}
\label{lem:bv-deleting-candidates-destr}
Bucklin voting is resistant to destructive control by deleting
candidates.
\end{lemma}

\begin{proofs}
Susceptibility holds by
  Lemma~\ref{lem:susceptible-candidate-control}.
To show the problem $\np$-hard, let $(C,V)$ be the election resulting
from a
{\sc Restricted Hitting Set} instance $(B,\mathcal{S},k)$ according to
Construction~\ref{con:bv-resistance-general-candidate-control}, and let
$c$ be the distinguished candidate.

We claim that $\mathcal{S}$ has a hitting set of size at most~$k$ if
and only if $c$ can be prevented from being a unique BV winner by
deleting at most $m-k$ candidates.

From left to right: Suppose $\mathcal{S}$ has a hitting set $B'$ of
size~$k$. Delete the $m-k$ candidates $B-B'$.  Now, both candidates
$c$ and $w$ have a strict majority on level~2, but
\begin{eqnarray*}
\scoresublevel{(\{c,d,w\} \cup B', V)}{2}{c} & = & 4n(k+1)+2(m-k)+9, \\
\scoresublevel{(\{c,d,w\} \cup B', V)}{2}{w} & = & 4n(k+1)+2(m-k)+10,
\end{eqnarray*}
so $w$ is the unique level~2 BV winner of this election.

From right to left: Suppose that $c$ can be prevented from being a
unique BV winner by deleting at most $m-k$ candidates.  Let $D'
\subseteq B\cup \{ d,w\} $ be the set of deleted candidates (so
$c\notin D'$) and $D = (C-D')-\{ c \}$.  It follows immediately from
Lemma~\ref{lem:bv-resistance-candidate-control} that $D = B'\cup \{
d,w \}$, where $B'$ is a hitting set for $\mathcal{S}$ of size at
most~$k$.~\end{proofs}

\begin{lemma}
\label{lem:bv-partition-candidates}
Bucklin voting is resistant to constructive and destructive control by
partition of candidates and by run-off partition of candidates (for
each in both tie-handling models, TE and~TP).
\end{lemma}

\begin{proofs}
  Susceptibility holds by
  Lemma~\ref{lem:susceptible-candidate-control}, so it remains to
  show $\np$-hardness.  For the constructive cases, map the given
  {\sc Restricted Hitting Set} instance $(B,\mathcal{S},k)$ to the
  election $(C,V)$ from
  Construction~\ref{con:bv-resistance-general-candidate-control} with
  distinguished candidate~$w$.

 We claim that $\mathcal{S}$ has a hitting set of size at most~$k$ if
 and only if $w$ can be made the unique BV winner by exerting control
 via any of our four control scenarios (partition of candidates with
 or without run-off, and for each in either tie-handling model, TE and
 TP).

 From left to right: Suppose $\mathcal{S}$ has a hitting set
 $B'\subseteq B$ of size $k$. Partition the set of candidates into the
 two subsets $C_1=B'\cup \{ c,d,w \}$ and $C_2=C-C_1$.  According to
 Lemma~\ref{lem:bv-resistance-candidate-control}, $w$ is the unique
 level~2 BV winner of subelection $(C_1, V) = (B' \cup \{c,d,w\},V)$.
 Since (no matter whether we have a run-off or not, and regardless of
 the tie-handling rule used) the opponents of $w$ in the final stage
 (if there are any opponents at all) each are candidates from
 $B$. Since $n>m$, $w$ has a majority in the final stage on the first
 level with a score of $4n(k+1)+9$.  Thus, $w$ is the unique BV winner
 of the resulting election.

From right to left: Suppose $w$ can be made the unique BV winner via
any of our four control scenarios.  Since $c$ is not a BV winner of
the election, there is a subset $D\subseteq B\cup \{ d,w\} $ of
candidates such that $c$ is not a unique BV winner of the election
$(D\cup \{ c\} ,V)$.  By Lemma~\ref{lem:bv-resistance-candidate-control},
there exists a hitting set for $\mathcal{S}$ of size at most~$k$.

For the four destructive cases, we simply change the roles of $c$ and
$w$ in the above argument.~\end{proofs}

Next we handle the one missing candidate-control case for Bucklin voting.
Our reduction in this case is from the $\np$-complete problem {\sc Hitting Set}
, which is defined as follows (see~\cite{gar-joh:b:int}):

\begin{desctight}

\item[Name:] {\sc Hitting Set}
.

\item[Instance:] A set $B=\{b_1,b_2,\ldots , b_m\}$, a collection
  $\mathcal{S} =\{S_1,S_2,\ldots , S_n\}$ of nonempty subsets
  $S_i\subseteq B$, and a positive integer $k\leq m$.

\item[Question:] Does $\mathcal{S}$ have a hitting set of size at most
$k$, i.e., is there a set $B'\subseteq B$ with $\|B'\|\leq k$ such that
for each $i$, $S_i \cap  B'\neq \emptyset$?  
\end{desctight}

\begin{lemma}
\label{lem:bv-deleting-candidates-constr}
Bucklin voting is resistant to constructive control by deleting candidates.
\end{lemma}

\begin{proofs}
  Susceptibility holds by
  Lemma~\ref{lem:susceptible-candidate-control}. To prove
  $\np$-hardness we give a reduction from {\sc Hitting Set}
  . Let
  $(B,\mathcal{S},k)$ be a {\sc Hitting Set}
   instance with
  $B=\{b_1,b_2,\ldots ,b_m\}$ a set, $\mathcal{S}=\{S_1,S_2,\ldots
  ,S_n\}$ a collection of nonempty subsets $S_i\subseteq B$, and
  $k\leq m$ a positive integer.
  Define the election $(C,V)$ with candidate set $C = B \cup C' \cup D
  \cup E \cup F \cup \{ w\},$ where $C'=\{ c_1, c_2, \ldots ,c_{k+1}
  \}$, $D=\{d_1,d_2,\ldots ,d_s\}$, $E=\{ e_1, e_2, \ldots , e_n\}$,
  $F = \{f_{1},\ldots,f_{n+k}\}$, $w$ is the distinguished candidate,
  and the number of candidates in $D$ is $s = \sum _{i=1}^n s_i$ with
  $s_i = n+k-\| S_i \|$, so $s = n^2+kn-\sum _{i=1}^n \| S_i\|$.
  For each $i$, $1\leq i \leq n$, let
  $D_i=\{d_{1 + \sum_{j=1}^{i-1} s_j}, \ldots , d_{\sum_{j=1}^{i} s_j} \}$.
  Define $V$ to be the following collection of $2(n+k+1)+1$ voters:
\begin{center}
\begin{tabular}{|c||l|c|l|}
 \hline
$\#$ & For each \ldots & number of voters &
 \multicolumn{1}{c|}{ranking of candidates}
 \\ \hline\hline
 1 & $i \in \{1, \ldots , n\}$ & $1$ & 
$
\begin{array}{c@{\ \ }c@{\ \ }c@{\ \ }c@{\ \ }c@{\ \ }c@{\ \ }c@{\ \ }c}
S_i & D_i & w & C' & E & (D-D_i) & (B-S_i)& F
\end{array}
$ \\ \hline
 2 & $j \in \{1, \ldots , k+1\}$ & $1$ & 
$
\begin{array}{c@{\ \ }c@{\ \ }c@{\ \ }c@{\ \ }c@{\ \ }c@{\ \ }c}
E & (C'-\{c_j\}) & c_j &  B & D & w & F
\end{array}
$ \\ \hline
 3 & & $k+1$ &
$
\begin{array}{c@{\ \ }c@{\ \ }c@{\ \ }c@{\ \ }c@{\ \ }c}
w & F & C' & E & B & D
\end{array}
$ \\ \hline
 4 & & $n$ &
$
\begin{array}{c@{\ \ }c@{\ \ }c@{\ \ }c@{\ \ }c@{\ \ }c}
C' & D & F & B & w & E
\end{array}
$ \\ \hline
 5 & & $1$ &
$
\begin{array}{c@{\ \ }c@{\ \ }c@{\ \ }c@{\ \ }c@{\ \ }c}
C' & w & D & F & E & B
\end{array}
$ \\ \hline
\end{tabular}
\end{center}

There is no unique BV winner in election $(C,V)$, since the
candidates in $C'$ and candidate $w$ are level $n+k+1$ BV winners.

We claim that $\mathcal{S}$ has a hitting set of size $k$ if and only
if $w$ can be made the unique BV winner by deleting at most $k$
candidates.

From left to right: Suppose $\mathcal{S}$ has a hitting set $B'$ of
size~$k$.  Delete the corresponding candidates.  Now, $w$ is the
unique level~$n+k$ BV winner of the resulting election.

From right to left: Suppose $w$ can be made the unique BV winner by
deleting at most $k$ candidates.  Since $k+1$ candidates other than
$w$ have a strict majority on level~$n+k+1$ in election $(C,V)$, after
deleting at most $k$ candidates, there is still at least one candidate
other than $w$ with a strict majority of approvals on level~$n+k+1$.
However, since $w$ was made the unique BV winner by deleting at most $k$
candidates, $w$ must be the unique BV winner on a level lower than or
equal to $n+k$.  This is possible only if in all $n$ votes of the
first voter group $w$ moves forward by at least one position.  This,
however, is possible only if $\mathcal{S}$ has a hitting set $B'$ of
size~$k$.~\end{proofs}

\subsection{Voter Control}
\label{sec:results:voter-control}

We now turn to voter control for Bucklin voting.
Our reductions are from the $\np$-complete problem 
 {\sc Exact Cover by Three-Sets} ({\sc X3C}, for
short), which is defined as follows (see~\cite{gar-joh:b:int}):

\begin{desctight}

\item[Name:] {\sc Exact Cover by Three-Sets} ({\sc X3C}).

\item[Instance:] A set $B = \{b_1, b_2, \ldots , b_{3m }\}$, $m\geq 1$, and a
collection $\mathcal{S} = \{S_1, S_2, \ldots , S_n\}$ of subsets $S_i
\seq B$ with $\|S_i\| = 3$ for each~$i$, $1 \leq i \leq n$.

\item[Question:] Is there a subcollection $\mathcal{S}' \seq
  \mathcal{S}$ such that each element of $B$ occurs in exactly one set
  in~$\mathcal{S}'$?
\end{desctight}

\begin{theorem}
\label{thm: bv-adding-voters-constr}
Bucklin voting is resistant to constructive control by adding voters,
by deleting voters, and by partition of voters in model TE and
model~TP, but is vulnerable to destructive control by adding and
by deleting voters.
\end{theorem}

\begin{lemma}
\label{lem:bv-adding-voters-constr}
Bucklin voting is resistant to constructive control by adding voters.
\end{lemma}

\begin{proofs}
Susceptibility holds by Lemma~\ref{lem:susceptible-voter-control}.
Let $(B,\mathcal{S})$ be an {\sc X3C} instance, where $B=\{b_1,b_2,\ldots
  ,b_{3m}\}$ is a set with $m>1$ and $\mathcal{S}=\{S_1,S_2,\ldots
  ,S_n\}$ is a collection of subsets $S_i\subseteq B$ with $\|S_i\|=3$
  for each $i$, $1 \leq i \leq n$.  (Note that {\sc X3C} is trivial to solve
  for $m=1$.)

Define the election $(C,V\cup V')$, where $C=B\cup \{w\} \cup D$ with
$D= \{d_1, \ldots ,d_{n(3m-4)} \}$ is the set of candidates, $w$ is
the distinguished candidate, and $V\cup V'$ is the following collection
of $n+m-2$ voters:
\begin{enumerate}
\item $V$ is the collection of $m-2$ registered voters of the form:
$
 \begin{array}{c@{\ \ }c@{\ \ }c@{\ \ }}
B &  D & w.
\end{array}
$
\item $V'$ is the collection of unregistered voters, where for each
  $i$, $1\leq i \leq n$, there is one voter of the form:
$
\begin{array}{c@{\ \ }c@{\ \ }c@{\ \ }c@{\ \ }c@{\ \ }}
D_i & S_i & w & (D-D_i) & (B-S_i),
\end{array}
$
where $D_i=\{d_{(i-1)(3m-4)+1}, \ldots , d_{i(3m-4)} \} $.
\end{enumerate}
Since $b_{1}\in B$ has a majority already on the first level, $w$ is
not a unique BV winner in $(C,V)$.

We claim that $\mathcal{S}$ has an exact cover for $B$ if and only if
$w$ can be made a unique BV winner by adding at most $m$ voters
from~$V'$.

From left to right: Suppose $\mathcal{S}$ contains an exact cover
for~$B$.  Let $V''$ contain the corresponding voters from~$V'$
(i.e., voters of the form
$\begin{array}{c@{\ \ }c@{\ \ }c@{\ \ }c@{\ \ }c@{\ \ }}
D_i & S_i & w & (D-D_i) & (B-S_i),
\end{array}$
for each $S_i$ in the exact cover)
and add $V''$ to the election.  It follows that 
$\scoresublevel{(C,V\cup V'')}{3m+1}{d_j}=m-1$ for all $d_j\in D$,
$\scoresublevel{(C,V\cup V'')}{3m+1}{b_j}=m-1$ for all $b_j\in B$, and
$\scoresublevel{(C,V\cup V'')}{3m+1}{w}=m$.  Thus, only $w$ has a
strict majority up to the $(3m+1)$st level and so $w$ is the unique
level $3m+1$ BV winner of the election.

From right to left: Let $V''\subseteq V'$ be such that $\|V''\|\leq m$
and $w$ is the unique winner of election $(C,V\cup V'')$.  Since $w$
must in particular beat every $b_j \in B$ up to the $(3m+1)$st level,
it follows that $\|V''\|=m$ and each $b_j\in B$ can gain only one
additional point.  Thus, the $m$ voters in $V''$ correspond to an
exact cover for~$B$.~\end{proofs}

\begin{lemma}
\label{lem:bv-deleting-voters-constr}
Bucklin voting is resistant to constructive control by deleting voters.
\end{lemma}

\begin{proofs}
Susceptibility holds by Lemma~\ref{lem:susceptible-voter-control}.
  Let $(B,\mathcal{S})$ be an {\sc X3C} instance as above.  Define the
  election $(C,V)$, where $C=B\cup \{c,w\} \cup D\cup F \cup G$ is the
  set of candidates with $D=\{ d_1, d_2, \ldots ,d_{3nm}\}$, $F = \{
  f_1, f_2, \ldots ,f_{3n(m-1)}\}$, and $G = \{ g_1, g_2, \ldots
  ,g_{3m(m-1)}\}$, and where $w$ is the distinguished candidate.
For each~$j$, $1\leq j \leq 3m$, define 
$\ell_j = \| \{S_i \in \mathcal{S} \condition b_j\in S_i \}\|$, and
for each $i, 1\leq i \leq n$, define\footnote{Note that 
$D_i=\emptyset$ if $\| B_i \| =3m$ and that $w$ is
always ranked at or later than the $(3m+1)$st position.
\label{foo:3m}}
\begin{eqnarray*}
B_i & = & \{b_j\in B \condition i\leq n-\ell _j\}, \\
D_i & = & \{d_{(i-1)3m+1}, \ldots , d_{3im-\| B_i \|} \}, \text{ and } \\
F_i & = & \{f_{(i-1)(3m-3)+1},\ldots,f_{i(3m-3)}\}.
\end{eqnarray*}
Also, for each~$k$, $1 \leq k \leq m-1$, define $G_k =
\{g_{3m(k-1)+1},\ldots,g_{3mk}\}$.  Let $V$ consist of the following
collection of $2n+m-1$ voters:
\begin{center}
\begin{tabular}{|c||l|c|l|}
 \hline
$\#$ & For each \ldots & number of &
 \multicolumn{1}{c|}{ranking of candidates}
 \\  &                 & voters &
 \\ \hline\hline
 1 & $i \in \{1, \ldots ,  n\}$ & $1$ &
$
\begin{array}{c@{\ \ }c@{\ \ }c@{\ \ }c@{\ \ }c@{\ \ }c@{\ \ }c@{\ \ }c@{\ \ }}
S_i & c & F_i & D & (B-S_i) & G & (F-F_i) & w
\end{array}
$ \\ \hline
 2 & $i \in \{1, \ldots ,  n\}$ & $1$ &
$
\begin{array}{c@{\ \ }c@{\ \ }c@{\ \ }c@{\ \ }c@{\ \ }c@{\ \ }c@{\ \ }c@{\ \ }}
B_i & D_i & w & F & (D-D_i) & (B-B_i) & G & c
\end{array}
$ \\ \hline
 3 & $k \in \{1, \ldots , m-1\}$ & $1$ &
$
\begin{array}{c@{\ \ }c@{\ \ }c@{\ \ }c@{\ \ }c@{\ \ }c@{\ \ }c@{\ \ }}
c & G_k & F & D & (G-G_k) & B & w
\end{array}
$ \\ \hline
\end{tabular}
\end{center}

Candidate $c$ is the unique level~$4$ BV winner in the election $(C,V)$.

We claim that $\mathcal{S}$ has an exact cover for $B$ if and only if
$w$ can be made the unique BV winner by deleting at most $m$ voters.

From left to right: Suppose $\mathcal{S}$ contains an exact cover for
$B$. By deleting the corresponding voters from the first voter group,
we have the following level~$3m+1$ scores in the resulting election
$(C,V')$:
\begin{eqnarray*}
\scoresublevel{(C,V')}{3m+1}{w} & = & n,\\
\scoresublevel{(C,V')}{3m+1}{b_i} & = & \scoresublevel{(C,V')}{3m+1}{c} = n-1 
\quad\text{for all $i$, $1\leq i \leq 3m$},\\
\scoresublevel{(C,V')}{3m+1}{d_j} & = & 1 \quad\text{or}\quad
  \scoresublevel{(C,V')}{3m+1}{d_j} = 0 \quad\text{for all $d_j\in D$, and} \\
\scoresublevel{(C,V')}{3m+1}{f_j} & = & \scoresublevel{(C,V')}{3m+1}{g_k} = 1
 \quad\text{for all $f_j \in F$ and all $g_k \in G$}.
\end{eqnarray*}
Since now there are $2n-1$ voters in the election, candidate $w$ is
the first candidate having a strict majority, so $w$ is the unique BV
winner of election $(C,V')$.

From right to left: Suppose $w$ can be made the unique BV winner by
deleting at most $m$ voters.  Since $w$ doesn't score any points on
any of the first $3m$ levels (see Footnote~\ref{foo:3m}), neither $c$
nor any of the $b_i$ can have a strict majority on any of these
levels.  In particular, candidate $c$ must have lost exactly $m$
points (up to the $(3m+1)$st level) after deletion, and this is
possible only if the $m$ deleted voters are all from the first or
third voter group.  On the other hand, each $b_i\in B$ must have lost
at least one point (up to the $(3m+1)$st level) after deletion, and
this is possible only if exactly $m$ voters were deleted from the
first voter group.  These $m$ voters correspond to an exact cover
for~$B$.~\end{proofs}

\begin{lemma}
\label{lem:bv-deleting-adding-voters-destructive}
Bucklin voting is vulnerable to destructive control by adding and
deleting voters.
\end{lemma}

\begin{proofs}
Susceptibility holds by Lemma~\ref{lem:susceptible-voter-control}.
  The polynomial-time algorithms that solve the two control problems
  for fallback voting~\cite{erd-rot:c:fallback-voting} can easily be
  adapted (e.g., by adjusting the input format from fallback elections
  to that for Bucklin elections) to work for Bucklin voting as well,
  as Bucklin voting is the special case of fallback voting where each
  voter approves of all candidates.~\end{proofs}

\begin{lemma}
\label{lem:bv-partition-voters-constructive}
Bucklin voting is resistant to constructive control by partition of
voters in both tie-handling models, TE and~TP.
\end{lemma}

\begin{proofs}
Susceptibility holds by Lemma~\ref{lem:susceptible-voter-control}.
  To show \np-hardness we reduce $\xthreec$ to our control
  problems.  Let $(B,\mathcal{S})$ be an {\sc X3C} instance with
  $B=\{b_1,b_2,\ldots ,b_{3m}\}$, $m\geq 1$, and a collection
  $\mathcal{S}=\{S_1,S_2,\ldots ,S_n\}$ of subsets $S_i\subseteq B$
  with $\|S_i\|=3$ for each $i$, $1 \leq i \leq n$. We define the
  election $(C,V)$, where $C = B\cup\{c,w,x\}\cup D \cup E \cup F \cup
  G$ is the set of candidates with $D = \{d_1,\ldots,d_{3nm}\}$, $E
  =\{e_1,\ldots,e_{(3m-1)(m+1)}\}$, $F
  =\{f_1,\ldots,f_{(3m+1)(m-1)}\}$, and $G =\{g_1,\ldots,g_{n(3m-3)}\}$,
  and where $w$ is the distinguished candidate. 
For each~$j$, $1\leq j \leq 3m$, define 
$\ell_j = \| \{S_i \in \mathcal{S} \condition b_j\in S_i \}\|$, and
for each $i, 1\leq i \leq n$, define
\begin{eqnarray*}
B_i & = & \{b_j\in B \condition i\leq n-\ell _j\}, \\
D_i & = & \{ d_{(i-1)3m+1}, \ldots , d_{3im-\| B_i \|} \}, \text{ and } \\
G_i & = & \{g_{(i-1)(3m-3)+1},\ldots,g_{i(3m-3)}\}.
\end{eqnarray*}
Also, for each~$k$, $1\leq k \leq m+1$, define $E_{k} =
\{e_{(3m-1)(k-1) +1},\dots,e_{(3m-1)k}\}$, and for each~$l$, $1\leq l
\leq m-1$, define $F_{l} = \{f_{(3m+1)(l-1)+1}, \dots, f_{(3m+1)l}\}$.
Let $V$ consist of the following $2n+2m$ voters:
\begin{center}
\begin{tabular}{|c||l|c|l|}
 \hline
$\#$ & For each \ldots & number of &
 \multicolumn{1}{c|}{ranking of candidates}
 \\
     &                 & voters &
 \\ \hline\hline
 1 & $i \in \{1, \ldots , n\}$ & $1$ &
$
\begin{array}{c@{\ \ }c@{\ \ }c@{\ \ }c@{\ \ }c@{\ \ }c@{\ \ }c@{\ \ }c@{\ \ }c@{\ \ }c@{\ \ }c@{\ \ }}
 c & S_i & G_i & (G-G_i) & F & D & E & (B-S_i) & w & x
\end{array}
$ \\ \hline
 2 & $i \in \{1, \ldots , n\}$ & $1$ &
$
\begin{array}{c@{\ \ }c@{\ \ }c@{\ \ }c@{\ \ }c@{\ \ }c@{\ \ }c@{\ \ }c@{\ \ }c@{\ \ }c@{\ \ }}
 B_i & D_i & w & G & E & (D-D_i) & F & (B-B_i) & c & x
\end{array}
$ \\ \hline
 3 & $k \in \{1, \ldots , m+1\}$ & $1$ &
$
\begin{array}{c@{\ \ }c@{\ \ }c@{\ \ }c@{\ \ }c@{\ \ }c@{\ \ }c@{\ \ }c@{\ \ }c@{\ \ }}
 x & c & E_k & F & (E-E_k) & G & D & B & w
\end{array}
$ \\ \hline
 4 & $l \in \{1, \ldots , m-1\}$ & $1$ &
$
\begin{array}{c@{\ \ }c@{\ \ }c@{\ \ }c@{\ \ }c@{\ \ }c@{\ \ }c@{\ \ }c@{\ \ }c@{\ \ }}
 F_l & c & (F-F_l) & G & D & E & B & w & x
\end{array}
$ \\ \hline
\end{tabular}
\end{center}
In this election, candidate $c$ is the unique level~$2$ BV winner with a
level~$2$ score of $n+m+1$.

We claim $\mathcal{S}$ has an exact cover $\mathcal{S}'$ for $B$ if
and only if $w$ can be made the unique BV winner of the resulting
election by partition of voters in both tie-handling models TE and TP.

From left to right: Suppose $\mathcal{S}$ has an exact cover
$\mathcal{S}'$ for $B$. Partition $V$ the following way. Let $V_{1}$
consist of:
\begin{itemize}
	\item the $m$ voters of the first group that correspond to 
the exact cover (i.e., those $m$ voters of the form 
	$	\begin{array}{c@{\ \ }c@{\ \ }c@{\ \ }c@{\ \ }c@{\ \ }c@{\ \ }c@{\ \ }c@{\ \ }c@{\ \ }c@{\ \ }c@{\ \ }}
		c & S_i & G_i & (G-G_i) & F & D & E & (B-S_i) & w & x
	\end{array}$ for which $S_{i}\in \mathcal{S}'$) and 
	
	\item the $m+1$ voters of the third group (i.e., all voters of the form 
	$\begin{array}{c@{\ \ }c@{\ \ }c@{\ \ }c@{\ \ }c@{\ \ }c@{\ \ }c@{\ \ }c@{\ \ }c@{\ \ }}
		x & c & E_k & F & (E-E_k) & G & D & B & w.
	\end{array}$
\end{itemize}
Let $V_{2} = V- V_1$. In subelection $(C,V_1)$, candidate $x$ is the
unique level~$1$ BV winner. In subelection $(C,V_2)$, candidate $w$ is
the first candidate who has a strict majority and moves on to the final
round of the election. Thus there are $w$ and $x$ in the final run-off,
which $w$ wins with a strict majority on the first level. Since both
subelections, $(C,V_1)$ and $(C,V_2)$, have unique BV winners, candidate
$w$ can be made the unique BV winner by partition of voters in both
tie-handling models, TE and {TP}.

From right to left: Suppose that $w$ can be made the unique BV winner by
exerting control by partition of voters. Let $(V_1,V_2)$ be such a
successful partition. Since $w$ wins the resulting two-stage election,
$w$ has to win at least one of the subelections (say, $w$ wins
$(C,V_1)$). If candidate $c$ participates in the final round, he or
she wins the election with a strict majority no later than on the
second level, no matter which other candidates move forward to the
final election. That means that in both subelections, $(C,V_1)$ and
$(C,V_2)$, $c$ must not be a BV winner.  Only in the second voter
group candidate $w$ (who has to be a BV winnner in $(C,V_1)$) gets
points higher than on the second-to-last level. So $w$ has to be a
level~$3m+1$ BV winner in $(C,V_1)$, which implies that there have
to be voters from the second voter group in $V_1$. Therefore, in
subelection $(C,V_2)$ only candidate $x$ can prevent $c$ from moving
forward to the final round.  Since $x$ is always placed behind $c$ in
all votes except those votes from the third voter group, $x$ has to be
a level~$1$ BV winner in $(C,V_2)$.  Since in $(C,V_1)$ it is not
possible that a candidate can tie with $w$ on the $(3m+1)$st level,
$w$ has to be the unique level~$3m+1$ BV winner in $(C,V_1)$. Thus
both elections $(C,V_1)$ and $(C,V_2)$ have unique BV winners and so
the construction works for both tie-handling models, TE and~{TP}.

It remains to show that $\mathcal{S}$ has an exact cover
$\mathcal{S}'$ for $B$. Since $w$ has to win $(C,V_1)$ with the votes
from the second voter group, not all voters from the first voter group
can be in $V_1$ (otherwise $c$ would have $n$ points already on the
first level). On the other hand, there can be at most $m$ voters from
the first voter group in $V_2$ because otherwise $x$ would not be a
level~$1$ BV winner in $(C,V_2)$.  To ensure that no candidate
contained in $B$ has the same score as $w$, namely $n$ points, and
gets these points on an earlier level than $w$ in $(C,V_1)$, there
have to be exactly $m$ voters from the first group in $V_2$ and these
voters correspond to an exact cover for $B$.~\end{proofs}

\section{Conclusions and Open Questions}
\label{sec:conclusions}

We have shown that Bucklin voting is resistant to all standard types
of candidate control and all standard types of constructive
control.  In total, it possesses at least 18 resistances to the 22 commonly
studied control types, it has at least two (and can have no more than
four) vulnerabilities, and two cases remain open: destructive control
by partition of voters in both tie-handling models, TE and~{TP}.  For
comparison, recall from Table~\ref{tab:summary-of-results} that, for
destructive control by partition of voters, approval voting is
vulnerable both in model TE and TP, SP-AV is vulnerable in model TE
but resistant in model TP, and fallback voting is resistant in model
TP and it is open whether fallback voting is vulnerable or resistant
to this control type in model~{TE}.

Only SP-AV and fallback voting are currently known to be resistant to
one more control type than Bucklin voting.  However, Bucklin voting is
arguably a simpler and more natural voting system; for example, unlike
SP-AV and fallback voting, it is a majority-consistent voting rule.

{
\bibliographystyle{alpha}

\bibliography{fv}
}

\clearpage

\appendix

\section{Some Results of \cite{hem-hem-rot:j:destructive-control} 
Used in Section~\ref{sec:results:susceptibility}}

\begin{theorem}[\cite{hem-hem-rot:j:destructive-control}]
 \label{thm:voiced-control}
 \begin{enumerate}
 \item If a voiced voting system is susceptible to destructive control
	by partition of voters (in model TE or TP), 
	then it is susceptible to destructive control by deleting voters.
 \item Each voiced voting system is susceptible to constructive control
	by deleting candidates.
 \item Each voiced voting system is susceptible to destructive control
	by adding candidates.\footnote{Following
Bartholdi et al.~\cite{bar-tov-tri:j:control},
Hemaspaandra et al.~\cite{hem-hem-rot:j:destructive-control}
considered only the case of control by adding a limited number of
candidates---the ``unlimited'' case was introduced only in (the conference
precursors of) \cite{fal-hem-hem-rot:j:llull-copeland-full-techreport}.
However, it is easy to see that all results about control by adding
candidates stated in
Theorems~\ref{thm:voiced-control},
\ref{thm:duality-constructive-destructive-control},
and~\ref{thm:susceptibility-implications} hold true in both the limited
and the unlimited case.}
\end{enumerate}
\end{theorem}

\begin{theorem}[\cite{hem-hem-rot:j:destructive-control}]
 \label{thm:duality-constructive-destructive-control}
 \begin{enumerate}
 \item A voting system is susceptible to constructive control by adding
candidates if and only if it is susceptible to destructive control by
deleting candidates.
 \item A voting system is susceptible to constructive control by deleting
candidates if and only if it is susceptible to destructive control by
adding candidates.
 \item A voting system is susceptible to constructive control by adding
voters if and only if it is susceptible to destructive control by
deleting voters.
 \item A voting system is susceptible to constructive control by deleting
voters if and only if it is susceptible to destructive control by
adding voters.
\end{enumerate}
\end{theorem}

\begin{theorem}[\cite{hem-hem-rot:j:destructive-control}]
 \label{thm:susceptibility-implications}
\begin{enumerate}
\item If a voting system is susceptible to constructive control by
partition of voters (in model TE or TP), then it is susceptible to
constructive control by deleting candidates.

\item If a voting system is susceptible to constructive control by
partition or run-off partition of candidates (in model TE or TP), then
it is susceptible to constructive control by deleting candidates.

\item If a voting system is susceptible to constructive control by
partition of voters in model TE, then it is susceptible to
constructive control by deleting voters.

\item If a voting system is susceptible to destructive control by
partition or run-off partition of candidates (in model TE or TP), then
it is susceptible to destructive control by deleting candidates.
\end{enumerate}
\end{theorem}

\end{document}